\documentclass{article}

\usepackage[a4paper,top=2cm,bottom=2cm,left=3cm,right=3cm,marginparwidth=1.75cm]{geometry}

\usepackage{amsmath,amsfonts}
\usepackage{graphicx}
\usepackage[colorlinks=true, allcolors=blue]{hyperref}
\usepackage[authoryear,round]{natbib}

\usepackage{authblk} 
\usepackage{setspace} 

\title{Spatial Heterogeneity in Climate Risk and Human Flourishing: An Exploration with Generative AI}

\author[1]{Stefano M. Iacus}
\author[2,3]{Haodong Qi}
\author[4]{Devika Jain}

\affil[1]{Institute for Quantitative Social Science, Harvard University, Cambridge, MA}
\affil[2]{Stockholm University Demography Unit, Stockholm, Sweden}
\affil[3]{Department of Global Political Studies, Malm\"o University, Malm\"o, Sweden}
\affil[4]{Center for Geographic Analysis, Harvard University, Cambridge, MA}

\date{}

\begin{document}

\maketitle

\begin{abstract}
Recent advances in Generative Artificial Intelligence (AI), particularly Large Language Models (LLMs), enable scalable extraction of spatial information from unstructured text and offer new methodological opportunities for studying climate geography. This study develops a spatial framework to examine how cumulative climate risk relates to multidimensional human flourishing across U.S. counties. High-resolution climate hazard indicators are integrated with a Human Flourishing Geographic Index (HFGI), an index derived from classification of 2.6 billion geotagged tweets using fine-tuned open-source Large Language Models (LLMs). These indicators are aggregated to the US county-level and mapped to a structural equation model to infer overall climate risk and human flourishing dimensions, including expressed well-being, meaning and purpose, social connectedness, psychological distress, physical condition, economic stability, religiosity, character and virtue, and institutional trust. The results reveal spatially heterogeneous associations between greater cumulative climate risk and lower levels of expressed human flourishing, with coherent spatial patterns corresponding to recurrent exposure to heat, flooding, wind, drought, and wildfire hazards. The study demonstrates how Generative AI can be combined with latent construct modeling for geographical analysis and for spatial knowledge extraction.
\end{abstract}


\section{Introduction}
Climate change is increasingly experienced as a spatially heterogeneous and compounding set of hazards that shape everyday life, livelihoods, and well-being across places. In the United States, the past decade has seen rising exposure to extreme heat, drought, wildfire, flooding, and hurricanes, with climate-related disasters occurring with unprecedented frequency and geographic reach. Geography scholars have long emphasized that environmental risks are fundamentally spatial phenomena, produced through interactions between physical hazards and place-based social, economic, and institutional conditions \citep{hewitt2014regions, Cutter2003, Blaikie2014} . From this perspective, climate change is not only an environmental process but also a spatially differentiated social process whose consequences vary systematically across geographic contexts.

A substantial body of research has documented the impacts of climate change on specific outcomes, such as morbidity \citep{haines2006climate, reid2016critical}, mental health \citep{burke2018higher, cunsolo2018ecological, lawrance2022ClimateMental}, economic losses \citep{botzen2019economic, boustan2020effect, roth2025local, smith2020billion}, or migration \citep{black2011migration,boas2019climate, cattaneo2019human, qi2023modelling, qi2026forecasting}. However, much of this work remains fragmented across domains. By contrast, geographers  have consistently emphasized multidimensional, relational approaches to well-being that situate individual outcomes within broader socio-spatial systems \citep{smith1994geography, massey2005space} . The concept of human flourishing provides a useful integrative framework for capturing these multiple dimensions, encompassing physical and mental health, meaning and purpose, social relationships, character and virtue, economic security, institutional trust, and spiritual life \citep{vanderweele2017promotion} . However, empiric linking of climate risk to such multidimensional outcomes at fine geographic resolution remains methodologically challenging.

Measurement has been a persistent obstacle. Survey-based approaches remain central to the study of subjective well-being, but are constrained by declining response rates, limited spatial coverage, and restricted temporal resolution, particularly in the context of rapidly evolving and spatially uneven climate hazards \citep{deaton2016context, nayak2019strengths}. Geographers have increasingly addressed these limitations through the use of spatially explicit secondary data and geo-computational methods that enable the integration of diverse data sources across scales \citep{Goodchild2007} . In this context, digital trace data have emerged as a valuable complement for capturing social processes and lived experience across space and time \citep{Elwood2011}. At the same time, some scholars have highlighted that such data are socially produced, unevenly distributed, and subject to systematic biases that must be explicitly addressed in spatial analysis \citep{Kwan2016, Elwood2011}.

Recent advances in Generative AI and climate analytics offer new opportunities to bridge this gap. Specifically, Large language models (LLMs) have expanded the methodological landscape of climate geography by enabling automated interpretation and synthesis of unstructured geo-referenced text data at unprecedented scale. Building on earlier methods in geocomputation and spatial analysis \citep{longley2015geographic}, emerging work in Generative AI demonstrates how machine learning and foundation models can support spatial reasoning, geographic knowledge extraction, and decision-making \cite{Janowicz2019, Li2024}. On the climate side, high-resolution hazard modeling now allows physical risks to be quantified consistently across multiple dimensions and geographies. On the social side, large-scale digital trace data provide an alternative lens on well-being \citep{iacusporro2021book, iacusporro2022significance}, on aspirations and intentions \citep{qi2025Google}, as well as on hate and negative attitudes towards minority groups \citep{arcila2024online}. Most importantly, such data can be processed in near real-time, and thus offer timely insights into various social phenomena. 

Leveraging these advances, this study develops a spatial framework to examine how cumulative climate risk relates to multidimensional human flourishing across U.S. counties. Relying on the Climate Risk and Resilience-adjusted Index (CRI) \citep{AlphaGeo_ClimateRisk2025} and the Human Flourishing Geographic Index (HFGI) \citep{HFGI}, we estimate a structural equation model (SEM) in which a latent climate risk construct (derived from the six observed hazard indicators) serves as an exogenous predictor of nine latent human flourishing dimensions: subjective well-being, meaning and purpose, character and virtue, social connectedness, psychological distress, physical condition, economic stability, religious and spiritual life, and institutional trust. By translating theoretically grounded survey constructs into spatially explicit indicators of expressed flourishing and linking them to a latent multi-hazard climate risk construct using structural equation modeling, we provide a spatially coherent assessment of how climate risk is associated with material, social, psychological, and existential dimensions of well-being.

By linking high-resolution climate risk data with large-scale, geographically explicit measures of human flourishing, this study makes three contributions. First, it advances a multidimensional and spatially explicit understanding of climate impacts that extends beyond health or economic loss to encompass social, moral, existential, and institutional domains. Second, it demonstrates how Generative AI can be used within geography as a reproducible method for spatial knowledge extraction from large-scale unstructured digital trace datasets. Third, it provides a spatially explicit empirical foundation for policy discussions about climate adaptation, resilience, and well-being, highlighting where climate risk may pose the greatest threats to human flourishing in the United States and beyond. Finally, it contributes to discussion on responsible Generative AI in Geography by foregrounding the ethical and epistemological implications of deploying Large Language Models in spatial analysis and climate adaptation research.

\section{Datasets and Methodology}

This study integrates high-resolution spatial climate data with large-scale, Generative AI-derived measures of expressed human flourishing to examine how cumulative climate risk relates to multidimensional well-being across U.S. counties. The analytical framework combines (i) Generative Artificial Intelligence for geographic knowledge extraction from unstructured text, (ii) spatial climate hazard indicators and (iii) latent variable modeling to capture underlying climate risk and human flourishing dimensions. Together, these components  links physical hazards, social media signals, and spatial inference within a unified spatial framework. 

Our analysis relied on two spatial data sets: (1) Climate Risk and Resilience-adjusted Index (CRI) \citep{AlphaGeo_ClimateRisk2025}, and (2) Human Flourishing Geographic Index (HFGI)  \citep{HFGI}, which were both aggregated at the US county-level.  The below sub-sections describe the datasets in detail.

\subsection{Generative-AI based Human Flourishing Geographic Index (HFGI)}

HFGI was inspired by Harvard’s Human Flourishing Program, aiming to capture expressions of human flourishing through the analysis of 2.6 billion tweets in the US from the Harvard CGA’s Geotweet Archive v2.0 \citep{geotweet2016}. These tweets are geolocated and enriched with census tract variables using the U.S. Census Bureau’s TIGER/Line Shapefiles (2020) \citep{tiger2020}. The analysis of tweets spans January 2013 to June 2023 and is performed by applying specifically fine-tuned open-source large language models (LLMs) at scale \citep{finetuning2024}. 

Global Flourishing Survey (GFS) has been the primary source for Harvard’s Human Flourishing Program to elicit individuals’ reflections on their lives, emotions, and attitudes \citep{gfs, vanderweele2017promotion}. However, declining survey response rates \cite{nayak2019strengths}, observer effects \cite{deaton2016context}, and high costs limit high-frequency measurement. In contrast, HFGI relies on information from the Internet and social networking platforms, whereby individuals spontaneously share reactions to events without prompting from researchers. Such an approach can serve as a barometer, revealing organic behavior and monitoring social trends at scale, such as happiness \cite{greyling2025happiness} and subjective well-being \cite{iacusporro2021book, iacusporro2022significance}.  It also minimizes  demand effects on respondents, and offers a complementary lens to the insights drawn from traditional surveys. 

To derive the measurements of human flourishing, we selected 46 questions from the GFS questionnaire and reformulated them into concise prompts for a large language model (LLM). In our analysis, a LLM is asked about whether a tweet is related to each of the selected questions and, if so,  give an answer to the related question on a three-level scale (low, medium, high). For example, as for the happiness questions: a label of ``high" means true happiness, whereas ``low" corresponds to sadness. 

Our textual analysis goes beyond classical sentiment analysis and simple polarity detection. Specifically, we fine-tuned the Llama 3.2 3B model \citep{llama32_3b}. The choice of a relatively small model
reflects a deliberate trade-off between classification accuracy and the scalability required to process
2.6 billion tweets. It is well established that, with appropriate fine-tuning, smaller models can
perform competitively with their larger counterparts \citep{finetuning2024}. Indeed, models within the same family (e.g., the Llama series) are typically trained on similar or even identical corpora. The number of parameters primarily determines the model's ability to generalize across tasks and respond to a broader range of inputs. Fine-tuning, in contrast, does not introduce new knowledge but rather helps the model
specialize and focus on a specific task. 

Another reason why we chose version 3.2 is that this model officially supports 8 languages (English, German, French, Italian, Portuguese, Hindi, Spanish, and Thai) and this is key for our application which aims at scaling further this work to the rest of the global geo Twitter archive. We also validate how fine-tuning works across languages. In a separate work, we found that fine-tuning helps the model to scale beyond the official eight languages that Llama pre-trained upon \citep{learningtopic}. Specifically, it enables the model to understand the task more than the language. This is due to the fact that the model already know how to map languages through their embedding space.

Overall, our fine-tuned LLM carries conceptual understanding of human flourishing dimensions. It also generalizes across new slang, orthographic variants, and evolving idioms without retraining. This semantic adaptability removes the need for “word-shift” calibration used in traditional natural language processing (NLP) pipelines and ensures consistency of interpretation over the 2013–2023 period and beyond.

To classify 2.6 billion tweets, we take advantage of several high performance computer clusters. We
used mainly NVIDIA A100/H100 gpus, running quantized fine-tuned Llama models for a total
amount of 1,045,048 hours. We also apply several check-pointing strategies and parallel distribution tricks to ensure the computing process can be resumed in the event of unexpected disruptions. Scalable database engines, like DuckDB \citep{duckdb2019} were used to store temporary calculations and execute vectorized computations on disk rather than in memory. 

This novel use of Generative AI enables scalable, context-aware geographic knowledge extraction while maintaining conceptual alignment with established well-being frameworks. Importantly, these indicators are interpreted as expressed human flourishing, publicly articulated signals embedded in digital traces, rather than direct measures of experienced or self-reported well-being.

\subsection{Spatial Climate Risk and Resilience-Adjusted Index (CRI)}

CRI is a database maintained by AlphaGeo, a company employing proprietary methods to downscale the latest Global Climate Models (GCMs) across six major hazard indicators (Heat Stress, Wildfire, Coast Flooding, Inland Flooding, Hurricane Wind and Drought) to the most granular coordinate level for every location on Earth. Each indicator represents a percentile score (0–100), where higher values indicate greater exposure to physical climate hazards.For this study, all CRI indicators are spatially aggregated to the U.S. county level, enabling consistent alignment with social indicators and downstream spatial modeling.. These indicators reflect the spatial rank of each geographic unit within the global distribution of hazard intensity, integrating both the frequency and potential severity of climate-related physical risks such as drought, heat stress, flooding, wind, and wildfire.  Rather than treating hazards independently, we conceptualize climate risk as a cumulative and overlapping spatial process, consistent emphasizing multi-hazard exposure and compounding risk.

\begin{figure}[h]
    \centering
    \includegraphics[width=1\linewidth]{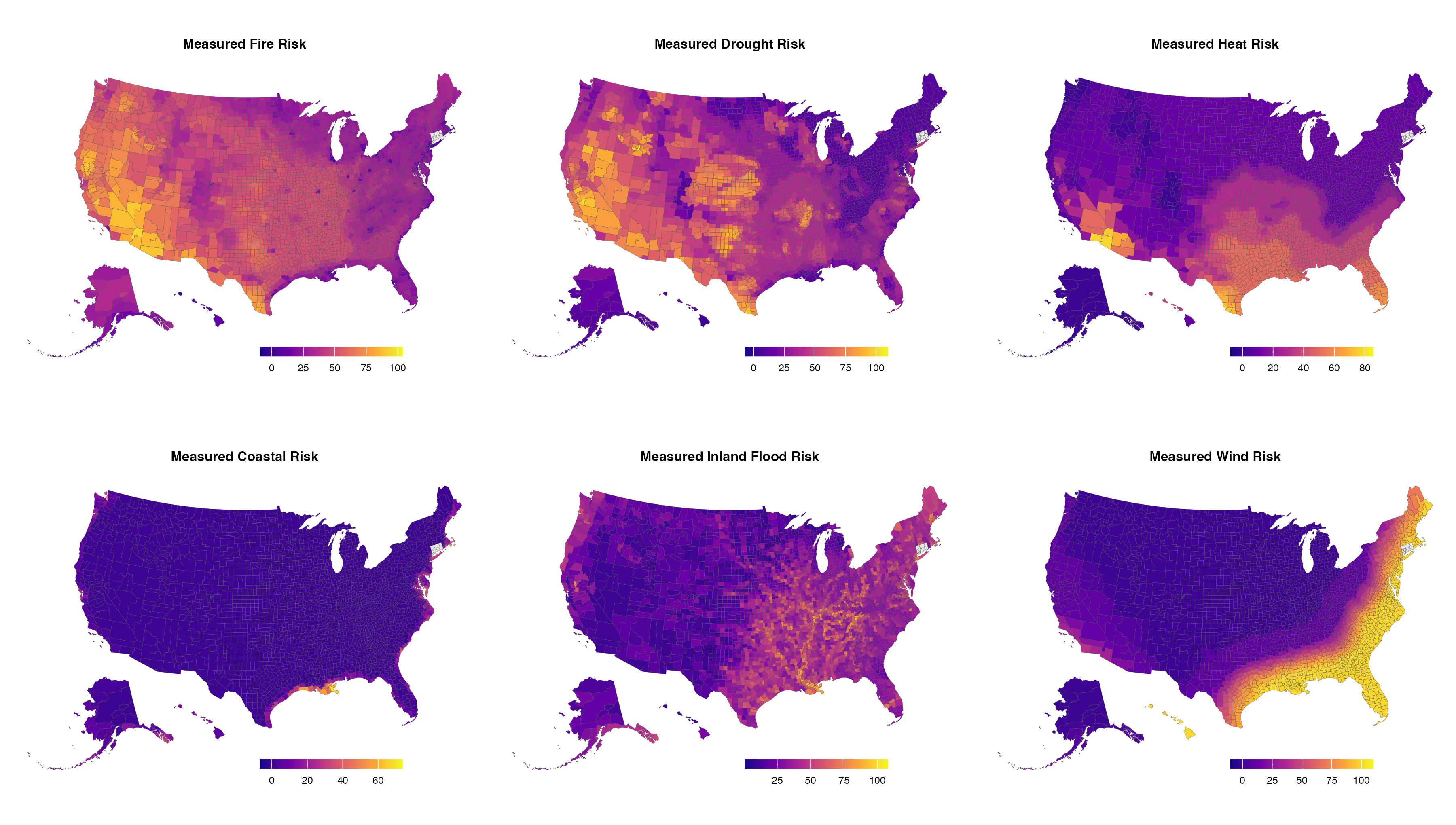}
     \caption{Measured Climate Risks}
    \label{fig:meas_clim}
\end{figure}

Figure~\ref{fig:meas_clim} depicts the spatial distribution and intensity of observed climatic hazards. Distinct regional patterns emerge across hazard types. Heat risk is highest across the South, Southwest, and parts of the Central Valley and Southern Plains. Wildfire risk is concentrated in the western United States, particularly California, the Pacific Northwest, and portions of the Mountain West. Drought risk is elevated across the Southwest, Southern Plains, and interior West, reflecting prolonged multi-year drought conditions during the 2010s and early 2020s. Coastal risk is concentrated along the Gulf Coast and Atlantic seaboard, while wind and inland flooding risks show elevated levels across the Midwest, Great Plains, and parts of the Southeast. In our analysis below, we use these observed indicators to form a latent climate risk surface that reflects the cumulative and overlapping exposure to climate-related hazards across U.S. counties during a decade characterized by increasing frequency and severity of extreme weather events. Table~\ref{tab:climate_risk_vars} in the Appendix summarizes the variables.

\subsection{Spatial Aggregation and Indicator Construction}

After classifying each tweet, we recoded the labels to numeric values: $-1$ if low, $+0.5$ if medium, and $+1$ if high, and $0$ if not-present. These numeric values are then used to build indicators through aggregation. Let $D_{i,k}$ be the numeric value for each geolocated tweet $i=1,...,2.6B$ and each question $k=1,...,46$, we first aggregate over census area $A$ for a given day $d$:
\begin{align}
    C^A_{k,d}=\sum_{i \in \tau^A_d} D_{i,k}
\end{align}
where, $\tau^A_d$ is a set of indexed tweets posted on day $d$ and geotagged by census area $A$.

We then aggregate $\tau^A_d$ to the county level and over all days during January 2013 - June 2023:
\begin{align}
    C^{j}_{k,t} = \sum_{d \in t} \sum_{A \in j} C^A_{k,d}
\end{align}
The indicator derived from each question $k$ is computed as:
\begin{equation}
    I^j_{k,t}=\dfrac{C^{j}_{k,t}}{n_{k,t,j}}
\end{equation}
where, $n_{k,t,j}$ is the number of tweets related to question $k$ during period $t$ in county $j$.

The resulting $I$'s are bounded $[-1, +1]$. It is important to note that these indicators are not the same as what captured by the Global Flourishing Survey, we therefore interpret them as the \textit{expressed} rather than \textit{realized} or \textit{experienced} human flourishing. Moreover, for 3 out of the 46 questions (related to the perceptions of immigration, corruption, and post-traumatic stress disorder), the LLM generated indicators contain little variations. They are therefore dropped in our analysis below. Tables~\ref{tab:flourishing_dimensions}-\ref{tab:flourishing_GFS} summarize the meaning of each of the 43 dimensions retained and their relaitonship with the original GFS survey. 

\begin{figure}[h]
    \centering
    \includegraphics[width=1\linewidth]{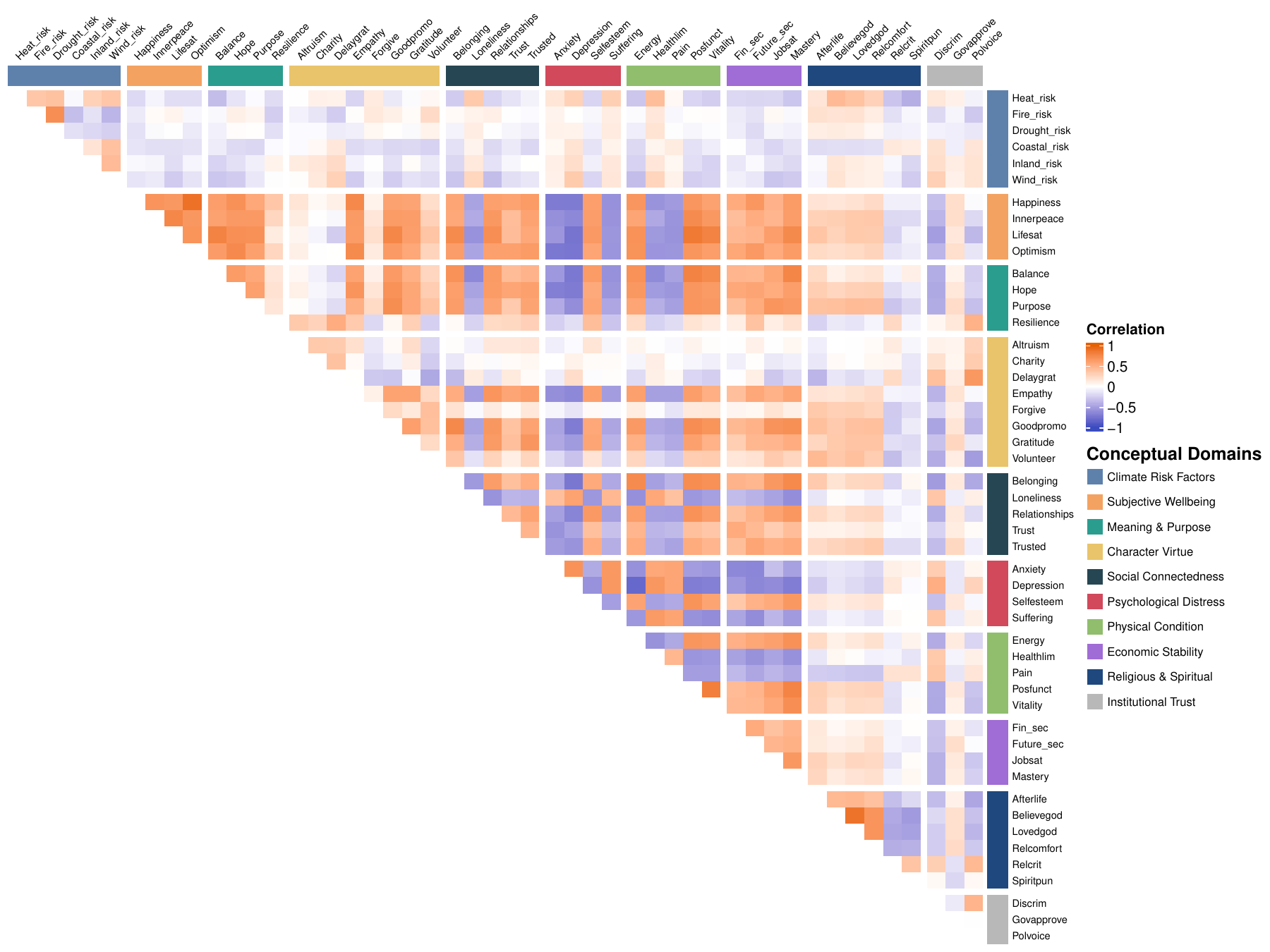}
     \caption{Correlations between the Human Flourishing indicators (HFGI) and the Climate Risk and Resilience (CRI) indexes.}
    \label{fig:corr_indicator}
\end{figure}

Figure~\ref{fig:corr_indicator} illustrates how our flourishing indicators co-vary with county-level climate risks. There are numerous strong correlations. For example, counties with higher heat risk tend to exhibit lower level of expressed \texttt{Happiness}, \texttt{Life satisfactions}, \texttt{Optimism}, \texttt{Belonging}, \texttt{Resilience}, \texttt{Empathy}, \texttt{Self-esteem}, \texttt{Energy}, \texttt{Vitality}, \texttt{Economic stability}, and higher level of \texttt{Loneliness}, \texttt{Depression}, \texttt{Suffering}, and \texttt{Health limitation} as well as stronger believe and love in god. 

This spatial aggregation strategy preserves geographic structure while enabling large-scale spatial inference. Indicators with insufficient spatial or temporal variation are excluded from subsequent analysis. The resulting dataset captures spatial heterogeneity in multiple dimensions of human flourishing and enables direct comparison with county-level climate risk.

\subsection{Structural Equation Modeling of Human Flourishing Dimensions}

To synthesize the indicators  $I^j_{k,t}$  and transform them into latent human flourishing dimensions, we estimate a  Structural Equation Modeling (SEM). The model consists of one exogenous construct (climate risk), and nine endogenous constructs capturing \texttt{Climate Risk Factors}, \texttt{Subjective Well-being}, \texttt{Meaning \& Purpose}, \texttt{Character Virtue}, \texttt{Social Connectedness}, \texttt{Psychological Distress}, \texttt{Physical Condition}, \texttt{Economic Stability}, \texttt{Religious \& Spiritual}, \texttt{Institutional Trust}. Climate risk was specified as the sole predictor of all endogenous latent variables. This setup allows us to examine how the nine human flourishing dimensions respond to climate risk exposures. 

The measurement model in our SEM is defined as:
\begin{align}
\mathbf{x} &= \Lambda_x \boldsymbol{\xi} + \boldsymbol{\delta}, \\
\mathbf{y} &= \Lambda_y \boldsymbol{\eta} + \boldsymbol{\varepsilon}
\end{align}
where, $\boldsymbol{\xi}$ denotes the exogenous latent variable (climate risk), $\boldsymbol{\eta}$ the vector of nine endogenous latent variables, $\mathbf{x}$ and $\mathbf{y}$ are the corresponding observed indicators, and $\boldsymbol{\delta}$ and $\boldsymbol{\varepsilon}$ are the associated error terms.

Observed indicators were linked to latent constructs through loading matrices $\Lambda_x \in \mathbb{R}^{6 \times 1}$ and $\Lambda_y \in \mathbb{R}^{43 \times 9}$. The measurement model followed a simple-structure specification, such that each indicator loaded on a single latent construct with no cross-loadings. Measurement error covariances were fixed to zero:
\begin{equation}
\mathrm{Cov}(\boldsymbol{\delta}) = \Theta_\delta, \qquad
\mathrm{Cov}(\boldsymbol{\varepsilon}) = \Theta_\varepsilon,
\end{equation}
where $\Theta_\delta$ and $\Theta_\varepsilon$ are diagonal matrices. Latent variables were scaled by fixing one factor loading per construct.

In the structural model of our SEM, all endogenous latent variables were regressed on climate risk:
\begin{equation}
\boldsymbol{\eta} = \Gamma \boldsymbol{\xi} + \boldsymbol{\zeta},
\end{equation}
where $\Gamma = (\beta_1,\ldots,\beta_9)^{\mathsf T}$ contains the structural coefficients, $\boldsymbol{\zeta}$ is the associated error term. No structural paths were specified among endogenous constructs. 

The variance of the exogenous latent variable was freely estimated: $\mathrm{Var}(\boldsymbol{\xi}) = \Phi$, and the residual variances of endogenous latent variables were freely estimated, whereas all residual covariances were constrained to zero: $\mathrm{Cov}(\boldsymbol{\zeta}) = \Psi = \mathrm{diag}(\psi_1,\ldots,\psi_9)$. Moreover, we assume no cross-loadings, no correlated measurement errors, no residual covariances among endogenous latent variables, and no structural relations among endogenous constructs beyond their dependence on climate risk. Given these constraints, the implied covariance matrix of the observed variables is given by:
\begin{equation}
\Sigma(\theta)
=
\begin{pmatrix}
\Lambda_x & 0 \\
0 & \Lambda_y
\end{pmatrix}
\begin{pmatrix}
\Phi & 0 \\
0 & \Gamma \Phi \Gamma^{\mathsf T} + \Psi
\end{pmatrix}
\begin{pmatrix}
\Lambda_x^{\mathsf T} & 0 \\
0 & \Lambda_y^{\mathsf T}
\end{pmatrix}
+
\begin{pmatrix}
\Theta_\delta & 0 \\
0 & \Theta_\varepsilon
\end{pmatrix}.
\end{equation}
SEM allows joint estimation of measurement and structural relationships while accounting for measurement error and latent spatial structures. All flourishing dimensions are regressed on latent climate risk, enabling systematic assessment of how cumulative hazard exposure relates to multiple facets of well-being across space.

\section{Results}

\subsection{Structural Parameter Estimates of Climate Risk and Human Flourishing} 

\begin{figure}[h]
    \centering
    \includegraphics[width=1\linewidth]{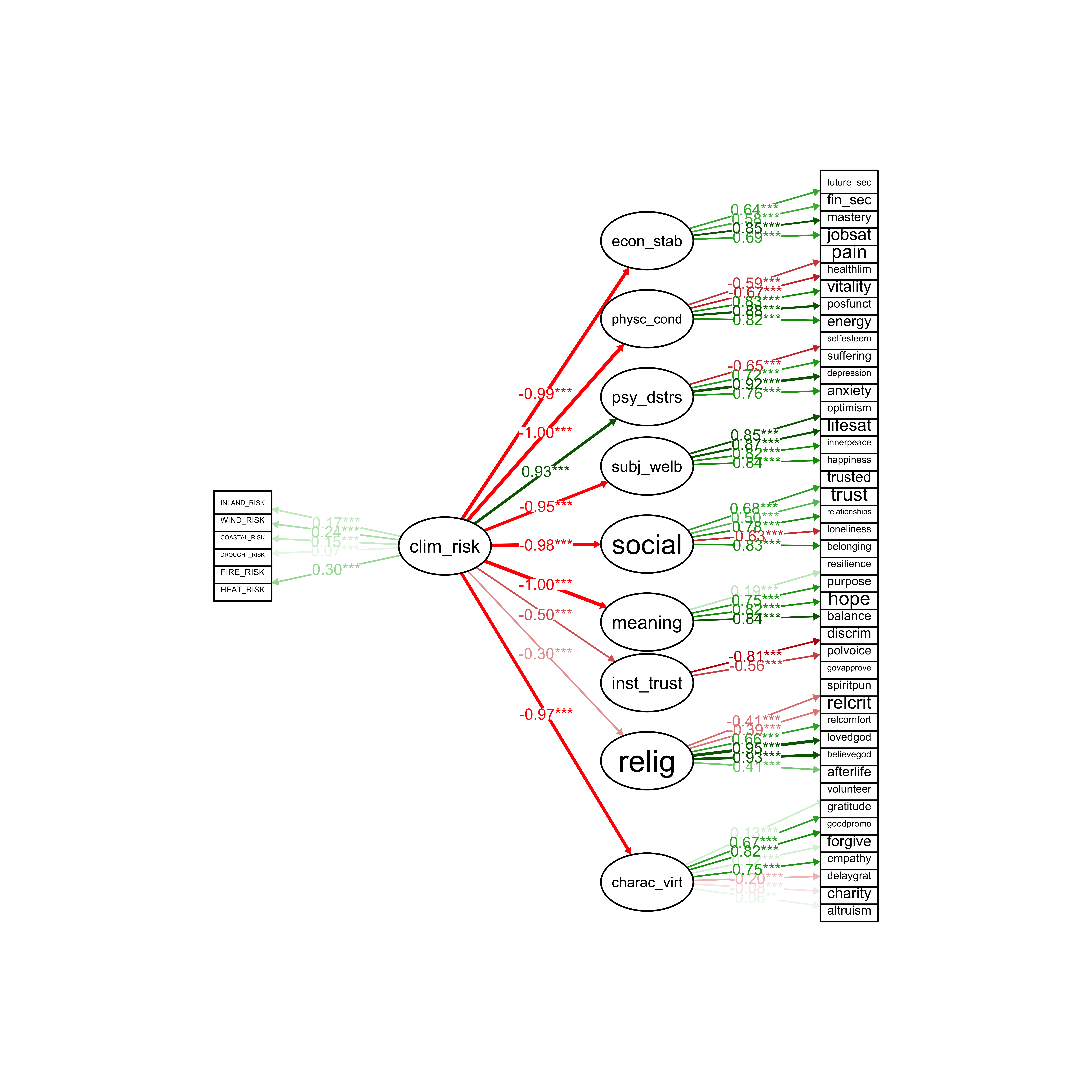}
     \caption{Estimates of Structural Equation Model. Note: *$p<.05$, **$p<.01$, ***$p<.001$. }
    \label{fig:sem_path}
\end{figure}

Figure~\ref{fig:sem_path} illustrates the parameter estimates of our SEM. All latent constructs were well identified, with moderate to strong standardized factor loadings. The factor loadings were statistically significant at $p< .05$, and their signs align with theoretical expectations. For example, \texttt{Future security}, \texttt{Financial security}, \texttt{Mastery of skills and ability}, \texttt{Job satisfaction} are positively associated with economic stability. Physical condition is negatively correlated with \texttt{Pain}, \texttt{Health limitations}, whereas positively associated with \texttt{Vitality}, \texttt{Positive function}, {Energy}.  Psychological distress increases with higher levels of expressed \texttt{Depression}, \texttt{Anxiety}, \texttt{Suffering}. Subjective well-being indicators showed consistently high loadings ($\lambda = 0.82–0.85$). Social connectedness grows strongly with \texttt{Trust}, \texttt{Belonging}, \texttt{Good relationships}, whereas deteriorates with \texttt{loneliness}. \texttt{Purpose}, \texttt{Hope}, \texttt{Balance} are strong predictors for meaning in life. Institutional trust declines with rises of \texttt{Discrimination}, \texttt{Political Voice}. Believe in and love for God are key determinants of religiosity. Collectively, the majority of 43 flourishing indicators support the construct validity of nine latent flourishing dimensions. In terms of climate, risks of \texttt{Heat}, \texttt{Wind}, \texttt{Inland flooding}, and \texttt{Costal erosion} exhibit relatively stronger loadings, suggesting that extreme high temperature, precipitation, as well as onsets of storms and/or hurricane are perceived as major climatic hazards in the US context.

The structural paths among latent variables imply significant effects of climate risk on human flourishing -- all paths between climate risk and latent flourishing dimensions are statistically significant at $p<0.05$. A higher level of perceived climate risk leads to more negative expression of Economic stability ($\beta= -0.99$), physical condition ($\beta= -1.00$), subjective well-being ($\beta= -0.95$), social connectedness ($\beta = -0.98$), meaning in life ($\beta= -1.00$), institutional trust ($\beta = -0.30$), religiosity ($\beta = -0.97$), and character virtues ($\beta = -0.97$). Meanwhile, it significantly heightens the level of psychological distress ($\beta=0.93$). 

Overall, the SEM estimates reveal a coherent and systematic structure underscoring the broad reach of climate risk perceptions across multiple domains of individual well-being. Specifically, when heat and wind risks elevate, population tend to feel deterioration in their material stability, physical and mental well-being, social relationships, as well as their trust and religious beliefs.

\subsection{Spatial Variation of Climate Risk and Human Flourishing }

Mapping the estimated latent climate risk scores reveals pronounced spatial heterogeneity across U.S. counties, while the spatial distribution of latent human flourishing dimensions exhibits varying degrees of correspondence with the geography of climate risk.

\begin{figure}[h]
    \centering
    \includegraphics[width=1\linewidth]{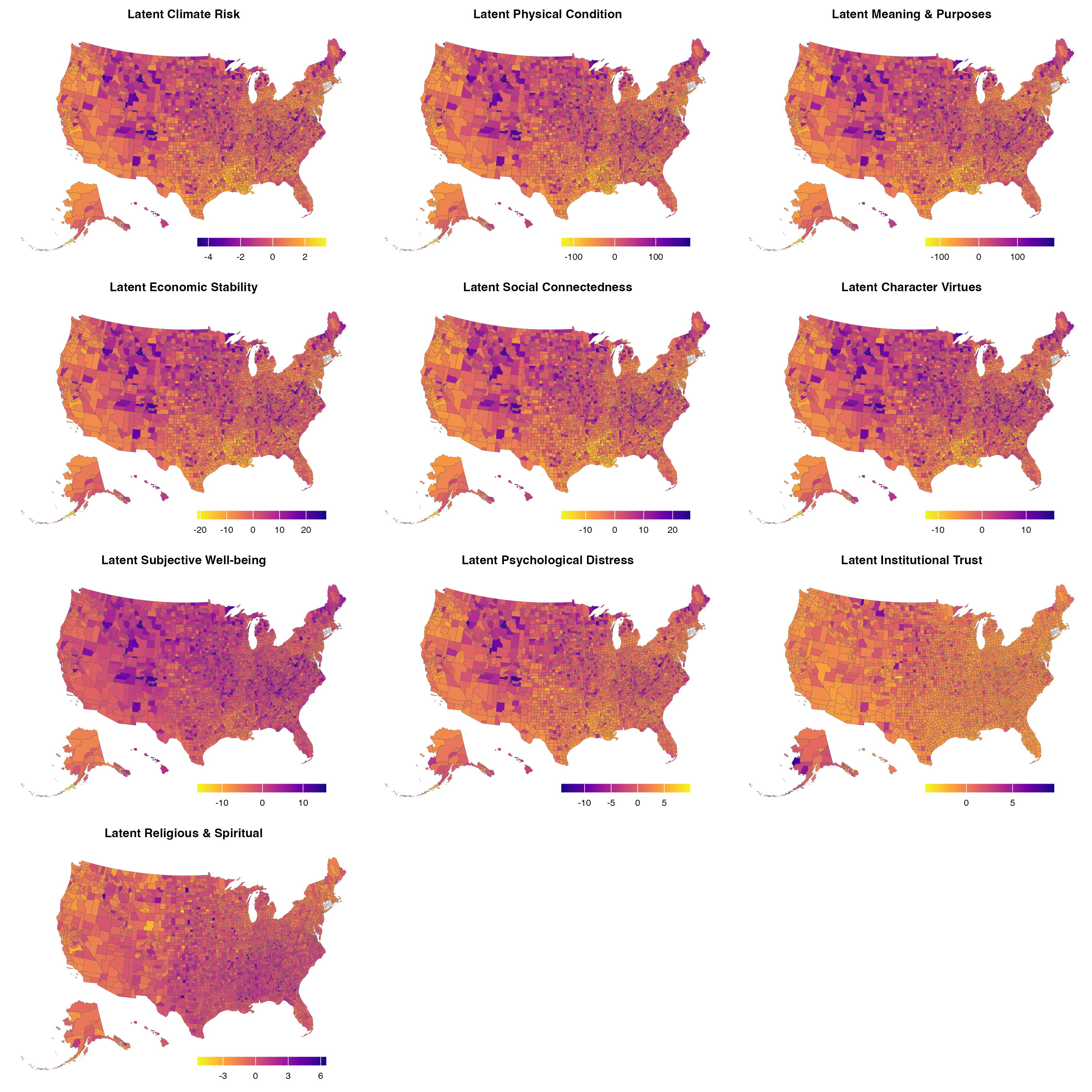}
     \caption{Spatial Variation of Latent Climate Risk and Human Flourishing Dimensions across US Counties, January 2013 - June 2023. Note: Panels are ordered in descending absolute standardized path from climate risk (cf. Figure~\ref{fig:sem_path}).}
    \label{fig:lat_var}
\end{figure}

Figure~\ref{fig:lat_var} maps county‑level scores for the latent climate risk factor and nine flourishing dimensions -- in the order of descending absolute standardized coefficients from latent climate risk in Figure~\ref{fig:sem_path}. Latent climate risk exhibits pronounced spatial heterogeneity -- concentrated in the Southeast and Gulf Coast, the Southwest and West Coast, and portions of the Midwest. We observe strong spatial clustering of high-risk values in the Southeast. These regions correspond closely to areas that experienced repeated exposure to hurricanes, extreme heat, drought, wildfires, and flooding between 2013 and 2023. Lower predicted risk is observed in parts of the Upper Midwest, Northern Plains, and New England. Thess patterns are consistent with national trends documented during the study period, including a steady rise in billion-dollar weather disasters, record-breaking wildfire seasons in the West, intensifying Atlantic hurricane activity, and widespread heat extremes \citep{smith2020billion,abatzoglou2016impact,knutson2020tropical}. 

The \texttt{physical condition} map shows broad swaths of lower vitality, energy, and functional health in the southern heat corridor, western regions with intense drought and wildfire, and major inland floodplains. In contrast, better physical health tends to cluster in portions of the Upper Midwest, Mountain West, and New England. These spatial patterns likely reflect the cumulative health consequences of repeated climate shocks, including disruptions to healthcare access, as well as chronic exposure to heat, air pollution from wildfire smoke, and disaster-related injuries \citep{haines2006climate, reid2016critical}.

The \texttt{meaning \& purpose} surface exhibits extensive troughs across high‑risk belts in the South, Southwest, and hazard‑exposed coastal zones. These troughs co‑locate with the latent climate risk map, indicating that counties facing cumulative heat and repeated exposures to flood and wind are also those where residents express lower balance, hope, and sense of purpose. This spatial co‑movement matches the largest absolute path in Figure~\ref{fig:sem_path}, suggesting that existential aspects of flourishing are among the most spatially sensitive to multi‑hazard environments, consistent with earlier research on climate-related loss, uncertainty, and erosion of future orientation \citep{cunsolo2018ecological}.

The expressed \texttt{economic stability} exhibit broadly similar spatial gradients: people tend to perceive economically worse-off along hurricane‑exposed coasts, riverine corridors, and the western heat/fire belt. These regional gradients—reflect the cumulative economic consequences of repeated hazard exposure, such as disruptions to employment, damages to  housing, infrastructure, and properties, and rising insurance and/or recovery costs \citep{boustan2020effect, botzen2019economic, roth2025local}. 

The \texttt{social connectedness} map reveals contiguous bands of lower belonging, higher loneliness, and weaker relationships in many of the same high‑risk clusters (coasts, flood corridors, drought interiors). The visual coherence of these depressions with the climate‑risk surface accords with the very large negative path shown in Figure~\ref{fig:sem_path}, indicating that recurrent disruption may limit ties and networks among people, due to, for example, displacement and forced migration.

The \texttt{character \& virtue} factor (empathy, forgiveness, gratitude, among others) shows clear declines across the high climate risk zones, implying that sustained environmental stress may erode prosocial and moral resources unevenly across regions. This pattern aligns with evidence that chronic stress and material insecurity can undermine cooperative norms and other-regarding behaviors \citep{haushofer2014psychology}.

\texttt{Subjective well‑being} (measured by happiness, life satisfaction, optimism, inner peace) is widely reduced across locations with high climate risk, particularly in the Gulf Coast, the hot‑dry West, and flood‑prone interiors. These high risk places also reveal elevated level of \texttt{psychological distress} (anxiety, depression, suffering, low self‑esteem). Such patterns are consistent with growing evidence documenting increased anxiety, depression, and stress symptoms associated with climate-related disasters, prolonged uncertainty, and repeated recovery demands \citep{burke2018higher,lawrance2022ClimateMental}. The temporal scope of the analysis includes several nationally salient events—such as major hurricanes, extreme wildfire seasons, and the COVID-19 pandemic—that may have amplified psychological vulnerability in climate-exposed communities \citep{pfefferbaum2020mental}.

The \texttt{Institutional trust} surface displays more diffuse spatial gradients that only partially overlap with climate risk. This diffuse pattern may reflect broader dissatisfaction with governmental and public institutions during recent years, including perceptions of inadequate disaster response, uneven recovery, pandemic management failures, rising costs of living, and widening economic inequality \citep{estadieu2025institutional}.

The spatial gradient of \texttt{Religious and Spiritual} co-locates the least with that of climate risk. Higher levels of religiosity are concentrated in the South and parts of the Midwest, consistent with long-standing regional cultural differences. Unlike economic, physical, and psychological outcomes, religion appears less uniformly shaped by climate exposure, suggesting that cultural and spiritual dimensions may respond differently to environmental stressors than material or health-related domains \citep{jenkins2018religion, haluza2014religion}.

Overall, our results demonstrate that cumulative climate risk is associated with spatially structured variation in multiple dimensions of human flourishing. The strongest associations occur in dimensions tied to material conditions, physical and psychological health, social relations, and existential orientation, while institutional and spiritual dimensions exhibit more complex spatial dynamics. These findings illustrate the importance of integrating Generative AI–derived indicators with latent spatial modeling to reveal coherent geographic patterns that are not readily observable from individual datasets alone. The results provide empirical evidence that multi-hazard climate exposure corresponds to systematic spatial variation in expressed well-being across U.S. counties, underscoring the importance of spatially explicit, multidimensional approaches in climate-related geographic analysis.

\section{Summary and Discussions}
Using a large language model that is fine-tuned based on  2.6 billion geo-located tweets in the US, we generated Human Flourishing Geographic Index (HFGI), containing 46 indicators. These indicators are inspired by Harvard's Human Flourishing Program to elicit individuals’ reflections on their lives, emotions, and attitudes. By aggregating these indicators to US county-level, we further explore how expressed human flourishing may respond to climate risk. 

The spatial findings documented in the article suggest that cumulative climate risk is associated with widespread reductions in multiple dimensions of human flourishing. Several mechanisms likely underlie these relationships. First, repeated exposure to extreme weather events imposes direct physical and economic burdens through injury, illness, infrastructure and property damages, and income and employment disruption. Over time, these shocks can compound, producing chronic stress, resource depletion, and diminished capacity for recovery. Second, climate hazards disrupt social and institutional systems by displacing populations, straining public services, and weakening trust in governance, thereby undermining social cohesion and institutional legitimacy. Third, prolonged exposure to environmental instability may erode existential and psychological resources by increasing uncertainty about the future, limiting perceived control, and diminishing meaning and purpose in daily life. Notably, the strong spatial alignment between climate risk and existential dimensions of flourishing suggests that climate change may affect not only material conditions but also individuals’ sense of balance, hope, and purpose. These findings extend prior work on climate-related mental health impacts by highlighting existential vulnerability as a potentially central pathway linking environmental risk to overall well-being.

The results underscore the need for climate adaptation and mitigation policies that address not only physical infrastructure and economic resilience but also social, psychological, and institutional dimensions of well-being. While material supports may be most critical in economically vulnerable regions, investments in healthcare access, mental health services, and community-based recovery programs may be particularly important in regions experiencing repeated shocks. Policies aimed at strengthening local institutions, improving transparency and efficiency in disaster response, and reducing displacement may play a critical role in buffering erosion in social connectedness, fostering social cohesion, meaning, and resilience, and ultimately sustaining flourishing under increasing climate stress. 

A key contribution of this work lies in demonstrating how Generative AI can be used as a method for geographic measurement rather than solely as an analytical aid. The use of fine-tuned large language models enables the translation of theoretically grounded social science constructs into spatially explicit indicators derived from unstructured digital trace data. When combined with structural equation modeling, this approach allows for the joint estimation of measurement and structural relationships while accounting for latent structure and measurement error. However, several limitations warrant consideration in the process. First, the analysis is observational and based on modeled associations; causal interpretations should be made cautiously. Second, predicted latent scores reflect average conditions over a multi-year period and may mask short-term dynamics associated with specific events, such as the COVID-19 pandemic. Third, although the latent climate risk construct captures multiple hazards, it cannot fully represent local exposure and context. Fourth, cultural and historical factors may shape regional patterns in ways not fully accounted for in the model. Finally, as with any study relying on aggregated spatial data, our results may be subject to the Modifiable Areal Unit Problem (MAUP). The relationships observed at the coarse spatial resolution of the county level may obscure local spatial variations or micro-scale heterogeneity present at the neighborhood or census-tract level. Utilizing multi-scalar analyses could verify if these spatial dependencies persist at finer granularity. Moreover, future research could integrate longitudinal designs, finer temporal resolution, and policy indicators to better identify causal mechanisms and assess the human costs of climate risk. Linking latent flourishing outcomes to migration, political behavior, or health trajectories may yield deeper insight into the societal consequences of accelerating climate risk.

Overall, our study illustrates how Generative AI, when embedded within established spatial principles and spatial modeling frameworks, can support new forms of geographic inquiry into complex socio-environmental processes. By linking multi-hazard climate risk to multidimensional human flourishing through a spatially explicit, Generative AI  based pipeline, the work contributes to ongoing efforts to expand the methodological foundations of geography and spatial data science in the era of Generative AI.

\section*{Data Availability}
Data and replication scripts are available at Harvard Dataverse: \href{https://doi.org/10.7910/DVN/BEBJML}{https://doi.org/10.7910/DVN/BEBJML}.
An interactive dashboard is available at: \href{https://ai-services.dataverse.org/r/climateRiskHFGI/}{https://ai-services.dataverse.org/r/climateRiskHFGI/}.


\section*{Acknowledgments}
We acknowledge the support of Harvard FAS Research Computing for providing most of the
computational resources, the NSF ACCESS program for providing initial computing bandwidth
to start the analysis and Kempner Institute for the Study of Natural and Artificial Intelligence to
allow us to use spare cycles of their GPU cluster. We thank CGA’s Xiaokang Fu for his assistance
in the data processing. We also thank Parag Khanna for providing the AlphaGeo AI dataset on
county-level climate risk indicators. Funding support from the Belmont Forum, the Swedish Research Council Vetenskapsrådet (grant agreement 2022–06012-3) and U.S. National Science Foundation (Award Number: 2310908) is gratefully acknowledged.

\bibliographystyle{apalike}
\bibliography{sample}

\clearpage

\section*{Appendix}
\setcounter{table}{0}
\setcounter{figure}{0}
\renewcommand{\thetable}{A\arabic{table}}
\renewcommand{\thefigure}{A\arabic{figure}}

\begin{table}[h]
\centering
\caption{List of the Human Flourishing Geographic Index dataset dimensions.}
\label{tab:flourishing_dimensions}
{
\small
\setlength{\parskip}{0pt}
\setstretch{1.0} 
\renewcommand{\arraystretch}{1.0}
\begin{tabular}{p{4cm} p{9.5cm}}
\hline
\textbf{Variable name} & \textbf{Description (plain English)} \\
\hline
\texttt{afterlife} & Life after death belief \\
\texttt{altruism} & Expressing altruism \\
\texttt{anxiety} & Anxiety \\
\texttt{balance} & Balance in the various aspects of own life \\
\texttt{believegod} & Belief in God \\
\texttt{belonging} & Belonging to society \\
\texttt{charity} & Charitable giving or helping \\
\texttt{delaygrat} & Expressing delayed gratification \\
\texttt{depression} & Depression \\
\texttt{discrim} & Perceiving discrimination \\
\texttt{empathy} & Expressing empathy \\
\texttt{energy} & Having energy \\
\texttt{fearfuture} & Fear of future \\
\texttt{future\_sec} & $ = -1\times$ \texttt{fearfuture} \\
\texttt{finworry} & Financial or material worry \\
\texttt{fin\_sec} & $ = -1\times$ \texttt{finworry}  \\
\texttt{forgive} & Seeking for forgiveness \\
\texttt{goodpromo} & Promoting good \\
\texttt{govapprove} & Expressing government approval \\
\texttt{gratitude} & Expressing gratitude \\
\texttt{happiness} & Happiness \\
\texttt{healthlim} & Describing health limitations \\
\texttt{hope} & Having hope \\
\texttt{innerpeace} & Peace with thoughts and feelings \\
\texttt{jobsat} & Expressing job satisfaction \\
\texttt{lifesat} & Life satisfaction \\
\texttt{lovedgod} & Feeling loved by God \\
\texttt{loneliness} & Loneliness \\
\texttt{mastery} & Mastery (ability or capability) \\
\texttt{optimism} & Expressing optimism \\
\texttt{pain} & Feeling pain \\
\texttt{polvoice} & Feeling having a political voice \\
\texttt{posfunct} & Positive functioning \\
\texttt{ptsd} & PTSD (post-traumatic stress disorder) \\
\texttt{purpose} & Purpose in life \\
\texttt{relationships} & Quality of relationships \\
\texttt{relcomfort} & Feeling religious comfort \\
\texttt{relcrit} & Religious criticism \\
\texttt{resilience} & Resilience \\
\texttt{selfesteem} & Self-esteem \\
\texttt{spiritpun} & Spiritual punishment \\
\texttt{suffering} & Suffering \\
\texttt{trust} & Expressing trust \\
\texttt{trusted} & Feeling trusted \\
\texttt{vitality} & Vitality \\
\texttt{volunteer} & Volunteering 
\end{tabular}
}
\end{table}

\begin{table}[ht]
\centering
\caption{Approximate mapping of some of the  flourishing dimensions used in the Human Flourishing Geographic Index dataset and the original questions in the GFS survey.}
\label{tab:flourishing_GFS}
{
\small
\setlength{\parskip}{0pt}
\setstretch{1.0} 
\renewcommand{\arraystretch}{1.5}
\begin{tabular}{p{2.5cm} p{12.5cm}}
\hline
\textbf{HFGI Var.} & \textbf{GFS survey question} \\
\hline
\texttt{afterlife} & Do you believe in life after death, or not?  \\
\texttt{anxiety} & Over the last 2 weeks, how often have you been bothered by the following problems? Feeling nervous, anxious or on edge; Not being able to stop or control worrying
 \\
\texttt{balance} &  In general, how often are the various aspects of your life in balance? \\
\texttt{believegod} &  Do you believe in one God, more than one god, an impersonal spiritual force, or none of these? 
 \\
\texttt{belonging} &   How would you describe your sense of belonging
to your local community?\\
\texttt{charity} &  In the past month, have you donated money to a charity?  \\
\texttt{delaygrat} & I am always able to give up some happiness now for greater happiness later. 
  \\
\texttt{depression} & Over the last 2 weeks, how often have you been bothered by the following problems? Little interest or pleasure in doing things; Feeling down, depressed or hopeless
 \\
\texttt{discrim} &  How often do you feel discriminated against because of any group you are a part of? This might include discrimination because of your religion, political affiliation, race, gender, social class, sexual orientation, or involvement in civic organizations or community groups. 
 \\
\texttt{fearfuture} &  Just your best guess, on which step do you think you will stand in the future, say about five years from now? [Worst possible - Best possible] \\
\texttt{finworry} &  How often do you worry about being able to meet normal monthly living expenses? 
How often do you worry about safety, food, or housing? 
 \\
\texttt{forgive} & How often have you forgiven those who have hurt you?  \\
\texttt{goodpromo} & I always act to promote good in all circumstances, even in difficult and challenging situations. 
  \\
\texttt{govapprove} &  How much do you approve or disapprove of the job performance of the national government of this country? 
 \\
\texttt{gratitude} & If I had to list everything that I felt grateful for, it would be a very long list.  \\
\texttt{happiness} &  In general, how happy or unhappy do you usually feel? \\
\texttt{healthlim} & Do you have any health problems that prevent you from doing any of the things people your age normally can do?  \\
\texttt{hope} &  Despite challenges, I always remain hopeful about the future. \\
\texttt{innerpeace} & In general, how often do you feel you are at peace with your thoughts and feelings? 
\\
\texttt{lifesat} & Overall, how satisfied are you with life as a whole these days?  \\
\texttt{lovedgod} & I feel loved or cared for by God, the main god I worship, or the spiritual force that guides my life.\\
\texttt{loneliness} & How often do you feel lonely? 
 \\
\texttt{mastery} & How often do you feel very capable in most things you do in life? \\
\texttt{optimism} & Overall, I expect more good things to happen to me than bad. 
  \\
\texttt{pain} &  How much bodily pain have you had during the past 4 weeks?\\
\texttt{polvoice} & Do you agree or disagree with the following statement? People like me have a say about what the government does. 
\\
\texttt{ptsd} &  Think about the biggest threat to life you've ever witnessed or experienced first-hand during your life. In the past month, how much have you been bothered by this experience? \\
\texttt{purpose} &  Overall, to what extent do you feel the things you do in your life are worthwhile? 
\\
\texttt{relationships} & My relationships are as satisfying as I would want them to be. 
 \\
\texttt{relcomfort} & I find strength or comfort in my religion
or spirituality. \\
\texttt{relcrit} & People in my religious community are critical of me or my lifestyle. 
 \\
\texttt{volunteer} &  In the past month, have you volunteered your time to an organization?\\
\end{tabular}
}
\end{table}

\begin{table*}[ht]
\centering
\caption{\textbf{Climate risk variables and descriptions.}
All risk variables are percentile scores (0–100) derived from the
\textit{Climate Risk and Resilience-adjusted Index} (2005--2100).}
\label{tab:climate_risk_vars}
\begin{tabular}{p{1.2cm} p{4.55cm} p{9cm}}
\hline
\textbf{Variable} & \textbf{Expanded Name} & \textbf{Description} \\
\hline
\texttt{FIRE} &
 Wildfire Risk Score \par (Resilience-adjusted)&
Composite percentile (0--100) measuring exposure to wildfire hazard, incorporating
the effectiveness of local fire monitoring systems and defense infrastructure.
A higher score indicates greater wildfire risk. \\[3pt]

\texttt{DROUGHT} &
 Drought Risk Score \par (Resilience-adjusted)&
Percentile-based drought exposure index integrating the capacity of local water supply
systems and emergency water infrastructure.
Higher values correspond to greater resilience-adjusted drought risk. \\[3pt]

\texttt{HEAT} &
Heat Stress Risk Score \par (Resilience-adjusted)&
Percentile rank reflecting resilience-adjusted exposure to heat stress,
incorporating Urban Heat Island effects such as building density and green cover. \\[3pt]

\texttt{INLAND} &
 Inland Flooding Risk Score\par (Resilience-adjusted) &
Composite percentile of inland flooding risk that includes surface porosity, flood
control measures, and defensive infrastructure against riverine inundation. \\[3pt]

\texttt{COASTAL} &
 Coastal Flooding Risk Score \par (Resilience-adjusted)&
Percentile-based index reflecting exposure to coastal flooding and sea-level rise,
adjusted for the presence of coastal flood defenses and protection infrastructure. \\[3pt]

\texttt{WIND} &
 Hurricane Wind Risk Score \par (Resilience-adjusted)&
Percentile risk score representing hurricane-related wind exposure, incorporating
building strength and local storm mitigation capacity. \\[3pt]
\end{tabular}
\end{table*}

\end{document}